\documentclass[sigplan,nonacm]{acmart}

\usepackage{amsmath}
\usepackage{graphicx}
\usepackage{xspace}
\usepackage{mdframed}
\usepackage{tikz}

\begin{document}

\title{\sys: Software-Defined Hardware Prefetching}

\author{Keisuke Kamahori}
\affiliation{%
  \institution{University of Washington}
  \country{}
}

\author{Neil Adit}
\affiliation{%
  \institution{Meta}
  \country{}
}

\author{Kan Zhu}
\affiliation{%
  \institution{University of Washington}
  \country{}
}

\author{Yuqi Mai}
\affiliation{%
  \institution{Cornell University}
  \country{}
}

\author{Victor Lee}
\affiliation{%
  \institution{SiFive}
  \country{}
}

\author{Heiner Litz}
\affiliation{%
  \institution{University of California, Santa Cruz}
  \country{}
}

\author{Chris Kennelly}
\affiliation{%
  \institution{Google}
  \country{}
}

\author{Snehasish Kumar}
\affiliation{%
  \institution{Google}
  \country{}
}

\author{Hanna Alam}
\affiliation{%
  \institution{Google}
  \country{}
}

\author{Milad Hashemi}
\affiliation{%
  \institution{Google}
  \country{}
}

\author{David Li}
\affiliation{%
  \institution{Google}
  \country{}
}

\author{Adrian Sampson}
\affiliation{%
  \institution{Cornell University}
  \country{}
}

\author{Baris Kasikci}
\affiliation{%
  \institution{University of Washington}
  \country{}
}

\author{Tipp Moseley}
\affiliation{%
  \institution{Google}
  \country{}
}

\author{Parthasarathy Ranganathan}
\affiliation{%
  \institution{Google}
  \country{}
}

\author{Akanksha Jain}
\affiliation{%
  \institution{Google}
  \country{}
}

\newcommand{\sys}{\emph{Themis}\xspace}
\newcommand{\mpkithresh}{3\xspace}
\newcommand{\lambdaname}{usefulness factor\xspace}
\newcommand{\ie}{\textit{i.e.},\xspace}
\newcommand{\eg}{\textit{e.g.},\xspace}
\newcommand*\mycircle[1]{\tikz[baseline=(char.base)]{\node[shape=circle,draw,inner sep=1pt] (char) {\footnotesize #1};}}
\newcommand{\pgheading}[1]{\textbf{#1.}}
\newcommand{\bench}[1]{\texttt{#1}}
\newcommand{\insight}[2]{
\begin{mdframed}[nobreak=true,
  skipabove=\topsep,
  skipbelow=\topsep,
  linewidth=0.5mm
  ]
  \textbf{Observation:} #1\\
  \textbf{Insight:} \emph{#2}
\end{mdframed}
}

\fancyhead{}
\renewcommand{\headrulewidth}{0pt}

\begin{abstract}
Data cache misses represent a significant portion of stall cycles in datacenter workloads. Hardware prefetchers that reduce such stalls by fetching data ahead of time have become increasingly sophisticated. However, to achieve high coverage, they have to prefetch aggressively, generating many inaccurate accesses that waste memory bandwidth. This is problematic in datacenter environments where memory bandwidth is a limited resource due to high multi-tenancy. 

We observe that for datacenter workloads, inaccurate prefetches can be effectively filtered on a data page granularity, without sacrificing prefetch coverage. However, storing per-page metadata about prefetch usefulness in hardware is costly, so we propose a novel hardware-software interface for data prefetching: The software directs the hardware on where to prefetch, and the hardware identifies and issues prefetches in the regions of interest.

We propose \sys, a profile-guided hardware prefetching solution that implements this new interface. \sys utilizes page-level hints stored in page-table entries to disable the prefetcher for certain data pages at runtime. \sys requires no binary or ISA changes and can be used to optimize processes without disrupting their execution. \sys is also orthogonal to existing works on prefetching and can be applied to optimize any hardware prefetcher. Our results show that \sys is able to achieve around $40\%$ reduction in useless prefetch requests, resulting in speedup for all the evaluated prefetchers for datacenter workloads, including $4.1\%$ for BOP, $3.1\%$ for SPP+PPF, and $1.4\%$ for Pythia.

\end{abstract}


\maketitle
\begingroup
\renewcommand{\thefootnote}{}%
\footnote{Corresponding authors: Keisuke Kamahori
  (\nolinkurl{kamahori@cs.washington.edu}),
  Baris Kasikci (\nolinkurl{baris@cs.washington.edu}),
  Akanksha Jain (\nolinkurl{avjain@google.com}).}%
\addtocounter{footnote}{-1}%
\endgroup

\section{Introduction}

Modern datacenter applications exhibit large memory footprints and complex memory access patterns, causing frequent backend stalls~\cite{ayers2019asmdb,ayers2020classifying,ayers2018memory,ferdman2012clearing}. For instance, multiple datacenter providers have reported that $20\%$--$25\%$ of fleet-wide CPU cycles are wasted on backend stalls in their datacenters~\cite{ayers2019asmdb,sriraman2020accelerometer}. Data prefetching~\cite{falsafi2014primer,smith1978sequential,michaud2016best,kim2016path,bakhshalipour2019bingo,bhatia2019perceptron,bera2019dspatch,bera2021pythia,ros2019berti,pakalapati2020bouquet} is an effective technique for reducing backend stalls, as it can predict future memory accesses to hide their long latencies.

Unfortunately, due to their complexity, datacenter workloads do not benefit from hardware prefetchers as much as conventional benchmarks such as SPEC2017~\cite{SPEC2017} (see \S\ref{sec:motivation}). In fact, recent work has shown that in bandwidth-constrained environments, \textit{disabling} hardware prefetchers altogether can sometimes improve overall datacenter performance~\cite{jain2024limoncello}. The key problem is the poor accuracy of hardware prefetchers for datacenter workloads, which when combined with limited per-core available memory bandwidth~\cite{jain2024limoncello,mutlu2022modern} can result in poor overall performance. In environments with constrained bandwidth, prefetchers must act conservatively: they must not waste bandwidth with inaccurate prefetches in exchange for marginal cache hit rate improvements. Hardware throttling mechanisms such as CLIP~\cite{panda2023clip} aim to reduce inaccuracy, but they are not effective for datacenter workloads due to their large instruction and data footprints.

Profile-guided optimization (PGO) has been a highly impactful technique that allows post-silicon adaptation of hardware predictors in the face of diverse and evolving datacenter workloads. For example, it has shown significant improvements for instruction prefetching~\cite{ayers2019asmdb}, branch prediction~\cite{khan2022whisper} and cache replacement policies~\cite{khan2021ripple} by providing (1) the ability to leverage vast amounts of profiling data, (2) application-specific customizations, and (3) low overheads and minimal hardware modifications. In this paper, we leverage the strengths of PGO to improve the effectiveness of hardware prefetchers for datacenter workloads.

\begin{figure}[t]
    \centering
    \includegraphics[width=\columnwidth]{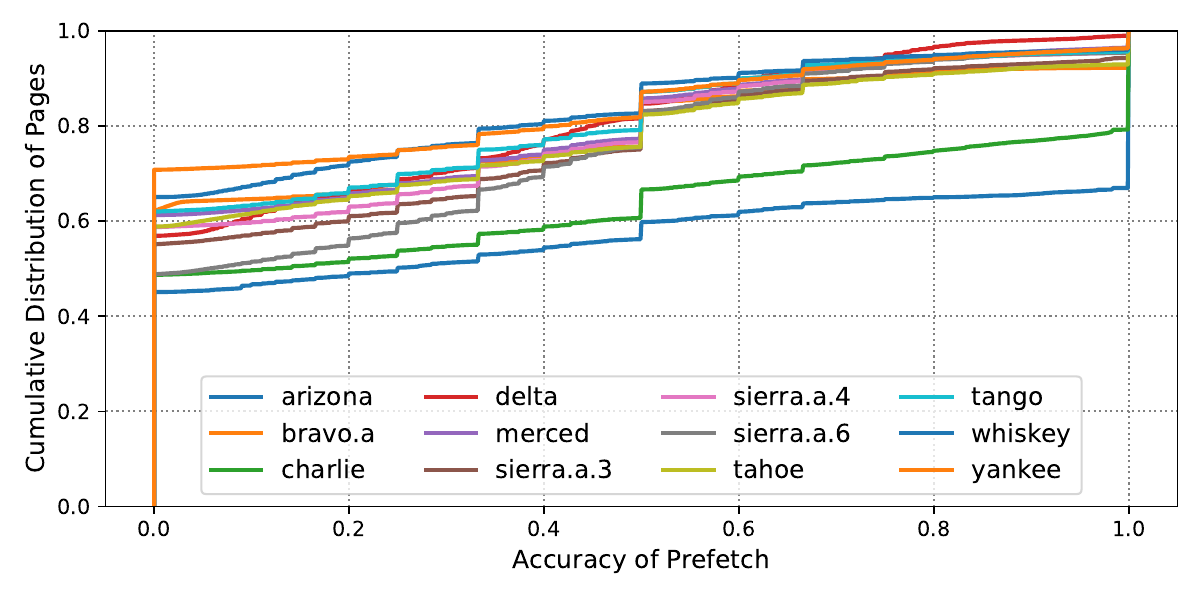}
    \caption{Cumulative distribution of the number of pages as a function of the prefetch accuracy. The x-axis shows the per-page accuracy of the prefetcher.}
    \label{fig:cdf-page-granularity}
\end{figure}

We perform a detailed datacenter workload study observing that prefetch efficiency can be significantly improved by enabling or disabling prefetchers at a data page granularity. Figure~\ref{fig:cdf-page-granularity} shows the cumulative distribution of pages as a function of prefetch accuracy for the Pythia~\cite{bera2021pythia} prefetcher for Google Workload Traces~\cite{Google_Workload_Traces_Version_2}. We can see that $50\%$--$75\%$ of observed pages have a prefetch accuracy of less than $20\%$ causing high bandwidth overheads for little gain (see \S\ref{sec:motivation} for more motivating data).

Tracking the effectiveness of prefetches on the page granularity is infeasible in hardware due to the large memory footprints. Throttling on the load-PC level is also ineffective due to the large number of loads and the fact that the same load can be sometimes effectively prefetched and sometimes not, depending on the page.

To address these challenges, we propose \sys, a mechanism implementing an effective division of labor between hardware and software.
We utilize PGO to determine which pages to prefetch and then utilize page-table attributes to communicate \emph{software directives} to the hardware.
Since page-table attributes are readily available in modern processors (\eg Arm's page-based hardware attributes~\cite{PBHA}) the required hardware modifications to implement \sys are minimal. Prefetchers only need to observe the added prefetch-enable bit contained within each TLB entry.
This division exploits the strengths of both domains: Software can capture large amounts of metadata effectively utilizing profiling, while the hardware prefetcher can observe dynamic memory accesses to generate prefetch candidates. By combining software’s accuracy with hardware’s coverage and timeliness, we enable post-silicon adaptation of prefetchers, allowing datacenters to optimize performance for evolving workloads without the need for costly hardware redesigns.

We evaluate \sys, implemented atop seven hardware prefetchers, using datacenter workloads from Google Workload Traces~\cite{Google_Workload_Traces_Version_2}, as well as benchmarks from SPEC2017~\cite{SPEC2017}, and GAP~\cite{beamer2015gap} suites.
For datacenter workloads, \sys improves instructions per cycle (IPC) upon all the seven prefetchers ($0.2\%$--$13.8\%$), including $4.1\%$ for BOP, $3.1\%$ for SPP+PPF, and $1.4\%$ for Pythia.
\sys reduces the number of useless prefetches on average by $40.6\%$ ($8.8\%$--$52.6\%$).
Furthermore, we observe that \sys can even increase the number of useful prefetches (coverage) by $5.7\%$ on average, despite only throttling prefetches for certain pages, since it mitigates cache pollution and directs hardware prefetchers to focus on data that is more likely to be beneficial, thereby enabling more efficient utilization of hardware resources.
\sys also achieves IPC improvement of $0.7\%$ ($0.1\%$--$2.2\%$) for SPEC2017 and $12.1\%$ ($1.3\%$--$35.4\%$) for GAP, outperforming existing throttling mechanisms~\cite{panda2023clip}.

Overall, we make the following contributions:
\begin{itemize}
    \item We show that state-of-the-art hardware prefetch throttlers are ineffective in datacenters due to coverage-accuracy trade-offs and a large instruction footprint.
    \item We propose \sys, a hardware-software co-designed solution that overcomes these issues by elevating fine-grained prefetch control to the software layer, such that the software can \emph{program} the hardware prefetcher at a page granularity based on runtime information. \sys can be applied to any hardware prefetcher.
    \item We evaluate \sys on seven state-of-the-art prefetchers across datacenter workloads. \sys improves IPC of all the evaluated prefetchers, including $4.1\%$ for BOP, $3.1\%$ for SPP+PPF, and $1.4\%$ for Pythia.
\end{itemize}

\section{Motivation}
\label{sec:motivation}
In this section, we investigate the performance of state-of-the-art hardware data prefetchers on datacenter applications and find that they are not very effective, and in fact, can be quite wasteful. We also discuss why existing hardware throttling solutions are insufficient for managing prefetchers in datacenter systems.

\subsection{Experimental Setup}
\label{sec:analysis-setup}
We perform all of our simulations using ChampSim~\cite{gober2022championship} (2022-01 release). We discuss our experimental setup in more detail in \S\ref{sec:eval}.
We characterize the following diverse set of existing hardware prefetchers on both Google Workload Traces~\cite{Google_Workload_Traces_Version_2} and traces of SPEC2017~\cite{SPEC2017}:
\begin{itemize}
    \item Nextline~\cite{smith1978sequential}
    \item Best Offset Prefetcher (BOP)~\cite{michaud2016best,michaud2015best}
    \item Signature Path Prefetcher (SPP)~\cite{kim2015lookahead,kim2016path}
    \item Bingo~\cite{bakhshalipour2019accurately,bakhshalipour2019bingo}
    \item SPP + Perceptron Prefetch Filtering (SPP+PPF)~\cite{bhatia2019enhancing,bhatia2019perceptron}
    \item Dual Spatial Pattern Prefetcher (DSPatch)~\cite{bera2019dspatch}
    \item Pythia~\cite{bera2021pythia}
\end{itemize}
We utilize the implementations of these prefetchers, as submitted by the corresponding authors to either the 2nd or the 3rd Data Prefetching Championship (DPC)~\cite{DPC2,DPC3} or published as open-source software. 
We only apply necessary modifications to accommodate the change in simulator versions.
For the following experiments, the prefetchers are placed at the L2 cache (the most common configuration among all the prefetchers listed above).
We measure IPC as well as statistics on \textit{useful} prefetches (prefetched data is accessed at least once before it is evicted from the cache) and \textit{useless} prefetches (prefetched data is never accessed before getting evicted from the cache).

\subsection{Why Do Existing Hardware Prefetchers Fall Short for Datacenter Workloads?}
\label{sec:analysis-pref-fall-short}

\begin{figure}[t]
    \centering
    \includegraphics[width=\columnwidth]{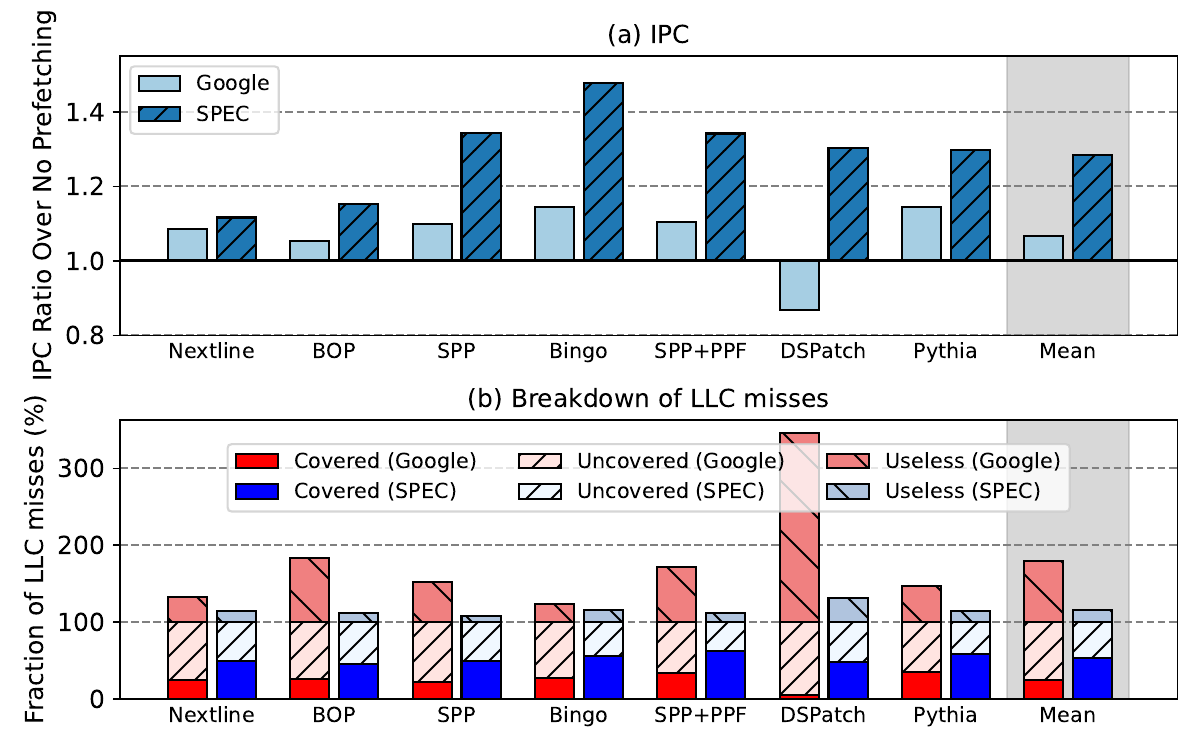}
    \caption{Comparison of (a) IPC and (b) breakdown of LLC misses of different prefetchers on Google traces~\cite{Google_Workload_Traces_Version_2} and SPEC2017~\cite{SPEC2017}.}
    \label{fig:analysis-prefetcher-performance}
\end{figure}

Figure~\ref{fig:analysis-prefetcher-performance}~(a) compares the IPC improvement of seven state-of-the-art hardware prefetchers for Google traces and SPEC2017, and we see that while most hardware prefetchers perform well for SPEC2017, they provide limited performance improvements for the Google traces. In particular, the geometric mean of speedup for the evaluated prefetchers is $6.8\%$ for Google traces, while $28.5\%$ for SPEC2017. With DSPatch, there is a slowdown of more than $13\%$. Pythia performs the best for Google traces, improving IPC by $14.5\%$. By contrast, the best-performing prefetcher for SPEC2017 (Bingo) improves performance by $47.8\%$.

Figure~\ref{fig:analysis-prefetcher-performance}~(b) sheds more insight into why existing hardware prefetchers perform poorly for datacenter workloads. The graph illustrates the ratio of (1) LLC misses covered by useful prefetches, (2) LLC misses not covered by prefetching, and (3) LLC misses due to useless prefetches for each prefetcher, all normalized to the number of misses without prefetching. 

We make two observations. First, the coverage (the fraction of misses covered by the prefetcher) is lower on Google traces. The average coverage on Google workloads is only $25.1\%$, with the best-performing prefetcher (Pythia) providing $35.8\%$ coverage. By contrast, the average coverage on SPEC is $53.0\%$, with the best-performing prefetcher (Bingo) providing $62.6\%$ coverage. 

Second, despite the lower coverage, hardware prefetchers exhibit a significant amount of useless prefetches for Google traces. On average, there are about $79.3\%$ of useless prefetches, even though the coverage is only $25.1\%$, which means the accuracy is only $24.0\% (= 25.1\% \div (25.1\% + 79.3\%))$. In contrast, SPEC2017 shows only $15.7\%$ of useless prefetches, and the accuracy is $77.1\%$.

In bandwidth-rich environments, prefetcher coverage is the primary indicator of its performance benefit: the higher the coverage, the better the overall performance. However, in bandwidth-constrained datacenter environments, low accuracy degrades performance by exacerbating bandwidth contention. 
In this paper, we focus on reducing the negative effects of bandwidth contention without compromising coverage.

\insight{
    Existing hardware prefetchers struggle to improve performance for datacenter applications due to low accuracy and wasteful prefetches.
}{
    It is critical to reduce useless prefetches and improve the accuracy.
}

\subsection{Why Do Existing Throttling Mechanisms Fall Short?}
\label{sec:analysis-throttle-fall-short}

Prefetchers typically have their built-in throttling mechanisms, with state-of-the-art solutions~\cite{panda2023clip,bhatia2019perceptron} even providing fine-grained prefetch filtering at the granularity of PCs. However, there are two fundamental reasons why these solutions are ineffective in the datacenter environment.
\begin{figure}[t]
    \centering
    \includegraphics[width=\columnwidth]{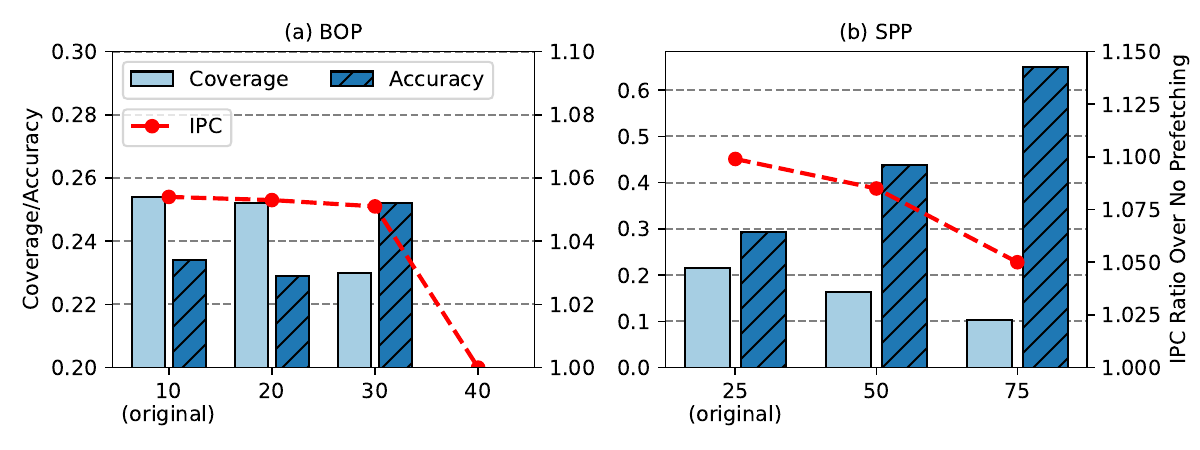}
    \caption{Coverage, accuracy, and IPC improvement over the original configurations of existing throttling mechanisms.}
    \label{fig:analysis-throttler}
\end{figure}

\pgheading{Coverage-Accuracy Trade-Off}
Throttling mechanisms inevitably introduce an accuracy-coverage trade-off, where a lot of coverage is compromised to achieve higher accuracy. This is problematic for datacenter workloads, where hardware prefetchers already have low coverage.

As a case study, Figure~\ref{fig:analysis-throttler} shows the coverage, accuracy, and IPC of BOP and SPP prefetchers, which have parameters for controlling the degree of throttling: \texttt{BADSCORE} for BOP and $T_P$ for SPP.
The default values are $10$ and $25$, respectively, and larger values indicate stronger throttling (fewer prefetches are issued).
Figure~\ref{fig:analysis-throttler} compares the performance with different values for these parameters, using Google traces.
For both prefetchers, stronger throttling improves accuracy, but degrades coverage and IPC, showing their ineffectiveness for datacenter workloads.

\pgheading{Large Instruction Footprint}
Moreover, tracking features such as program counters and page addresses within limited hardware budgets is challenging for datacenter workloads because they have large instruction and data footprints. For example, our evaluated datacenter workloads have more than $4\times$ larger instruction footprints than \bench{perlbench}, the largest among SPEC2017 workloads, and these footprints are constantly growing~\cite{kanev2015profiling}. Using hardware resources to track prefetching behavior at a fine granularity~\cite{bhatia2019perceptron,panda2023clip} can quickly require hundreds of kilobytes of hardware resources. 

\begin{figure}[t]
    \centering
    \includegraphics[width=\columnwidth]{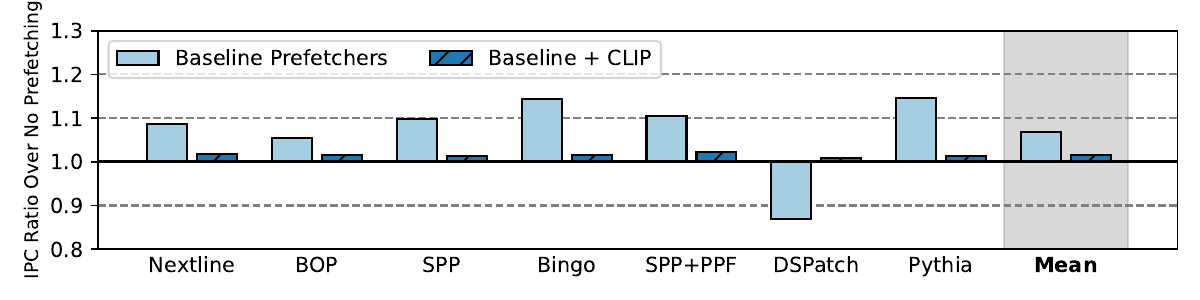}
    \caption{Comparison of IPC with and without CLIP for different prefetchers for Google traces.}
    \label{fig:analysis-clip}
\end{figure}

As a case study, CLIP~\cite{panda2023clip} is a state-of-the-art throttling technique for hardware prefetchers that blocks prefetches based on load instruction criticality (\ie whether a cache miss stalls the reorder buffer) and prefetch accuracy, using a buffer indexed by instruction addresses. Figure~\ref{fig:analysis-clip} shows that for Google traces, CLIP degrades performance for all prefetchers except for DSPatch. In datacenter applications, large instruction footprints cause over $67\%$ of prefetches to be blocked because the triggering instruction is absent from the buffer, leading to ineffective throttling. This data demonstrates that a large instruction footprint prevents the existing method from working effectively.

\insight{
    Existing throttling mechanisms are insufficient in datacenter environments, due to trade-offs of coverage and accuracy, and large instruction footprints.
}{
    A throttling technique that improves accuracy while maintaining coverage and requiring few hardware resources will be more effective for datacenters.
}

\subsection{How Do We Optimize Hardware Prefetchers for Datacenters?}
\label{sec:how-to-optimize}
By examining the behavior of hardware prefetchers for datacenter workloads, we observe significant variance in the usefulness of prefetching for different data addresses. 
Figure~\ref{fig:cdf-page-granularity} analyzes the impact of pages on prefetch accuracy for the Google traces, and we see that pages are a good discriminator for prefetch accuracy. In particular, we track the cumulative distributions of the number of pages (y-axis) as a function of the accuracy of prefetches issued to cache lines that reside on a page (x-axis). 
The figure illustrates the significant variance in prefetcher effectiveness across different memory regions. 
For instance, the prefetch accuracy is below $20\%$ for $50\%$--$75\%$ of the pages. 
If we disable prefetchers for the pages with the bottom $50\%$ accuracy, we can reduce useless prefetches with almost no loss of coverage.
Therefore, there exist ample opportunities for increasing the \emph{selectivity} of prefetchers based on page-level information.

To exploit this opportunity, neither hardware-only nor software-only methods work effectively. 
On the one hand, hardware-only methods suffer from the storage limitations as described earlier and will be shown later (\S\ref{sec:eval-sensitivity}).
On the other hand, while software prefetching is not constrained by hardware resource limitations, it lacks access to low-level hardware telemetry, which is necessary for page-level prefetch control.
Hence, we need \emph{hardware-software collaboration} to overcome these limitations. 

We utilize the profile-guided optimization (PGO) technique~\cite{khan2022whisper,zhang2022ocolos,zhang2024rpg2,jamilan2022apt} for hardware-software collaboration, enabling post-silicon adaptation of hardware prefetchers. PGO provides the best of both worlds: it combines timely prefetch generation in hardware, with the unconstrained metadata storage of software mechanisms.
Moreover, PGO allows the flexibility to optimize for evolving workloads without requiring hardware modifications. Since hardware design cycles span years, tuning prefetchers in advance for all workloads is impractical; PGO enables effective large-scale resource management.

From an implementation perspective, the page-table mechanism represents an excellent communication channel between hardware and software.
Modern CPUs already expose page-table attributes (\eg Arm’s page‑based hardware attributes~\cite{PBHA}) that can be manipulated by software, enabling \sys to be implemented with minimal hardware modifications.

\insight{
    Useless prefetches tend to be clustered in a select subset of pages, and it requires collaboration between hardware and software to exploit this opportunity.
}{
PGO represents a powerful technique to control and optimize prefetchers with minimal hardware modifications.
}

\section{Design of \sys}
\label{sec:design}
Our key insight is that hardware prefetchers' usefulness varies significantly across different memory regions. We leverage this insight to design \sys, a profile-driven throttling approach that introduces minimal hardware overheads. 

\subsection{Overview}
\sys controls hardware prefetchers at the granularity of the data page based on profile information. For pages where the prefetcher is not expected to issue useful prefetches, \sys disables the prefetcher. The hardware prefetcher is neither trained nor does it make predictions within those pages. Otherwise, \sys does not apply any modifications to existing prefetchers.

To determine which pages have prefetching enabled or disabled, \sys leverages runtime profile information. The profiling logic to obtain page-level accuracy information is implemented in software as a lightweight kernel module, and it requires minimal extensions to Cache Performance Monitor Units (PMUs) at the hardware level to obtain the necessary telemetry data. The obtained per-page {\em prefetching hints} are stored as page-based attributes within page-table entries (PTEs) and are replicated in the processor's TLB. For example, Arm's page-based hardware attributes (PBHA)~\cite{PBHA} expands PTEs and TLB entries with a $4$-bit field for software-managed implementation-defined attributes. 

This scheme has several advantages. First, it achieves hardware-software communication without requiring any ISA changes. Second, unlike many other PGO techniques~\cite{panchenko2021lightning,ottoni2021hhvm}, it does not require recompilation of the target process binary. Instead it can be applied to a running process without re-execution, as modifying the PTE entries only requires a relatively lightweight context switch into the kernel. The key insight enabling these advantages is that page-table attributes are an effective way to communicate performance hints from the software to the hardware at a fine granularity.

\subsection{Usage Model}
\label{sec:usage}
\begin{figure}[t]
    \centering
    \includegraphics[width=\columnwidth]{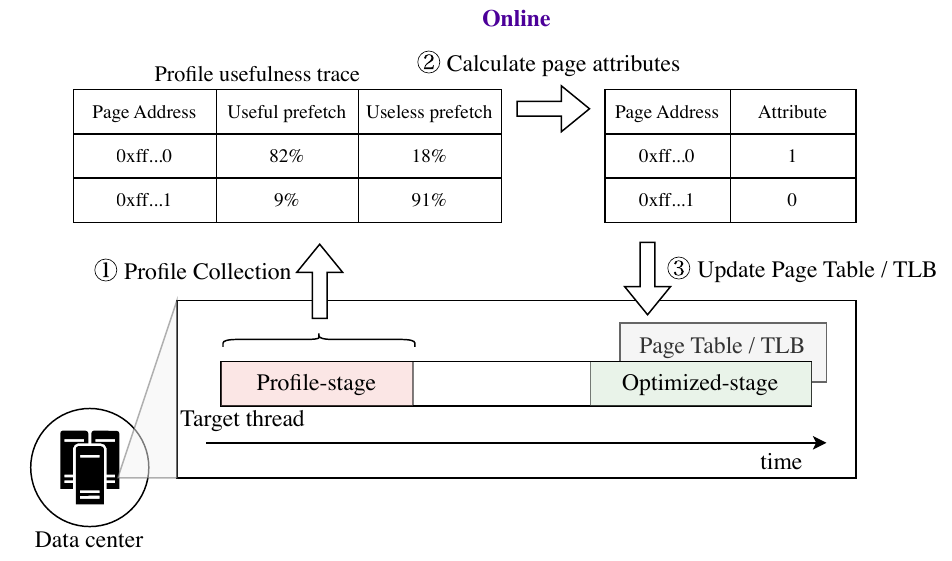}
    \caption{Usage model of \sys.}
    \label{fig:design-usage-model}
\end{figure}

We now show the end-to-end usage flow of \sys, before diving deeper into the profiling hardware, kernel module, and page-table attributes. 
Figure~\ref{fig:design-usage-model} shows the overview of the usage model of \sys. 
In step \mycircle{1}, \sys collects the prefetcher usefulness profile of applications at runtime. \S\ref{sec:design-microarch} gives the details of how prefetch usefulness profiles are collected.

Based on the profile data, in step \mycircle{2}, the kernel module in \sys calculates the page attributes from the ratio of the number of useful and useless prefetches. We use the criterion of Equation~\ref{eq:condition} to decide whether to disable the prefetcher for a given page:
\begin{equation}
    \lambda \times \text{\#useful} < \text{\#useless} \label{eq:condition}
\end{equation}
Here, $\lambda$ is a parameter named \emph{\lambdaname}, which is introduced to account for the different effects of useful and useless prefetches.
As the value of $\lambda$ decreases, the throttling increases, causing fewer prefetches to be emitted.
We empirically find $\lambda = 4$ to work best. 
More detail about the value of $\lambda$ is discussed in \S\ref{sec:eval-sensitivity}.

After that, in step \mycircle{3}, the calculated page attributes are injected into the process' page-table. Details of how the page-table attributes are injected and updated cheaply using existing APIs is discussed in the following.

\subsection{System Design}
\label{sec:design-system}

\begin{figure}[t]
    \centering
    \includegraphics[width=\columnwidth]{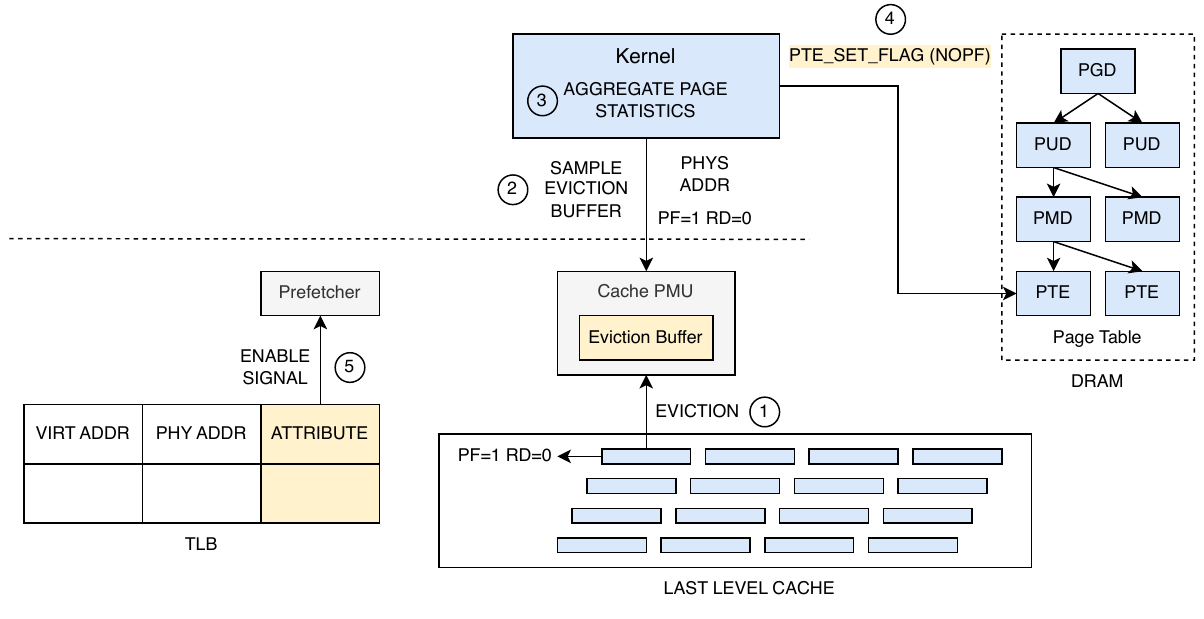}
    \caption{Microarchitecture and profiling mechanisms of \sys.}
    \label{fig:design-uarch}
\end{figure}

Figure~\ref{fig:design-uarch} shows the system components for \sys, including the microarchitectural modifications required for monitoring prefetch accuracy and the software logic needed to compute and update the hints. The profiling logic for obtaining page-level accuracy information is run in software as a kernel module. As detailed below, at the hardware level, only minor changes in the Cache PMU, a standard CPU component that monitors cache-related events, is required. All remaining data processing is handled within the kernel\footnote{The only privileged operation that needs to be performed in the kernel is the page-table walk for setting page attribute bits. Performance counter analysis can in principle also be implemented in user space, assuming the necessary PMU counter access permissions.}. 

\pgheading{Profiling Flow}
During the profiling phase, the following steps are performed (numbered in Figure~\ref{fig:design-uarch}):

\begin{enumerate}
    \item Cache evictions for useless prefetches are buffered in a sample buffer in the Cache PMU unit. The buffer stores physical addresses. Useless prefetches are identified if the prefetch bit is set in the tag when the line is evicted (the bit is cleared if the line was reused by a demand access).
    \item During the profiling phase, the kernel module periodically samples this buffer.
    \item The kernel module aggregates per-page accuracy statistics and updates the PTE attribute bits.
    \item At the end of the profile, the kernel updates the PTE attributes for sampled pages using \texttt{pte\_set\_flag} function~\cite{linux_pte_set_flags} in the kernel. The hints are percolated to the TLB \emph{lazily} the next time the TLB entry is updated for the page due to a TLB miss. This means that updating hints does not necessitate a TLB shootdown.
    \item The prefetcher is enabled/disabled based on the attribute in the TLB. 
\end{enumerate}

\pgheading{Microarchitectural Modifications}
\label{sec:design-microarch}
Our proposed design introduces two minor microarchitectural changes.

First, to support page-level profiling, we augment the Cache PMU with a small FIFO buffer designed to track the physical addresses of cache evictions that were prefetched but not used. Consequently, when a useless prefetch is detected, the PMU not only increments event counters such as \texttt{L2\_LINES\_OUT.USELESS\_HWPF}~\cite{IntelSDM3A} but also logs the evicted address into the FIFO buffer. This mechanism closely resembles Last Branch Record (LBR)-based sampling, which records the last N data addresses, and is commonly used for PGO techniques due to its low overhead~\cite{khan2022whisper}.

Second, to signal the prefetchers, the page-table and TLB entries are expanded with a field for page attributes. For \sys, we only use $1$ bit for each page. The TLBs for modern Arm processors are already augmented with similar attributes~\cite{PBHA}, and these bits are propagated through the memory subsystem with each memory operation. The page attribute is fed to a hardware prefetcher as an enable signal. If the prefetcher is disabled by the page attribute, it does not observe the new address from the CPU. 

\pgheading{Software Modifications}
\label{sec:design-kernel}
Most of the software modifications are to support data collection and hint generation during the profiling phase. A lightweight kernel module executes these functions. 

To collect per-page accuracy profiles, the kernel module samples the eviction buffer periodically. The overhead of retrieving this data is small as it is very similar to LBR-based sampling~\cite{khan2022whisper} (the only difference here is that we record data addresses in the Cache PMUs, not the processor pipeline).

The sampled addresses are used to compute the accuracy profiles at a page granularity as discussed in \S\ref{sec:usage}. Note that the addresses sampled by the Cache PMU are physical addresses, so the kernel module can only infer the physical page addresses. Fortunately, the Linux kernel maintains metadata indexed by physical page addresses~\cite{linux_physical_page} and hence can easily provide a reverse mapping to the corresponding virtual address~\cite{linux_reverse_mapping}. These utilities are used to associate page-level accuracy data with the corresponding virtual pages. The PTEs for virtual pages are retrieved via a software page walk, and the corresponding attribute field is updated using the \texttt{pte\_set\_flag} method. 

Note that updating performance attributes in the PTE does not require the kernel to initiate TLB shootdowns or cache maintenance operations since these are performance hints and not correctness-related flags.
In \sys shootdowns are not required since the page-table structure remains unchanged and hints can be communicated to the TLB with a delay. In particular, \sys communicates the hints \emph{lazily} as the hardware page walker fetches new translations into the TLB.

\pgheading{Prefetcher Throttling}
To implement the actual prefetch throttling, minimal hardware modifications are required. Hints are stored within TLB entries and are attached to every load/store packet after TLB lookup (this is already implemented in Arm processors where PBHA bits are propagated through the cache subsystem with each load/store packet). The hardware prefetcher observing loads and stores simply reads the hint bits and then enables or disables its training and prediction logic for the corresponding page.
Note that if L1 prefetchers are trained on virtual addresses, they may not have the hint available in time to pause training, but the predicted prefetch can be readily dropped because the prefetched entry needs to be translated through the TLB before it can be issued.

\subsection{Overhead Discussion}
\label{sec:design-overhead}

The main overhead of \sys stems from the profiling mechanism and the modification of the page-table attribute bits~\cite{PBHA}. As we will show, both overheads are low. 

First, the profiling needs to be performed infrequently which reduces the overall penalty, as we will quantify in \S\ref{sec:eval-overhead}.
Intuitively, profiling does not need to happen frequently as the prefetchability of a page is unlikely to change after its allocation. Prefetchability is inherently dependent on the data structure residing on the page which rarely changes. For example, memory regions storing arrays are likely to be prefetch-friendly, and those storing pointer-based accesses are likely to be prefetch-unfriendly~\cite{jamilan2022apt,ayers2020classifying}. Thus, the hints need to be updated roughly at the same cadence as pages are allocated and deallocated, which is not very frequent.

Second, the cost of updating the PTE attributes is around $1\,\mu\mathrm{s}$ per bit on average, based on the kernel module we implemented to traverse the page-table for reading and writing the PTE access bit.
Therefore, the overhead is only on the order of milliseconds for thousands of pages.
Although this may seem non-negligible, it is easily amortized over the extended execution times typical for datacenter applications that run for days or months. \S\ref{sec:eval-overhead} provides more supporting data.

\begin{figure*}[!htb]
    \centering
    \includegraphics[width=\linewidth]{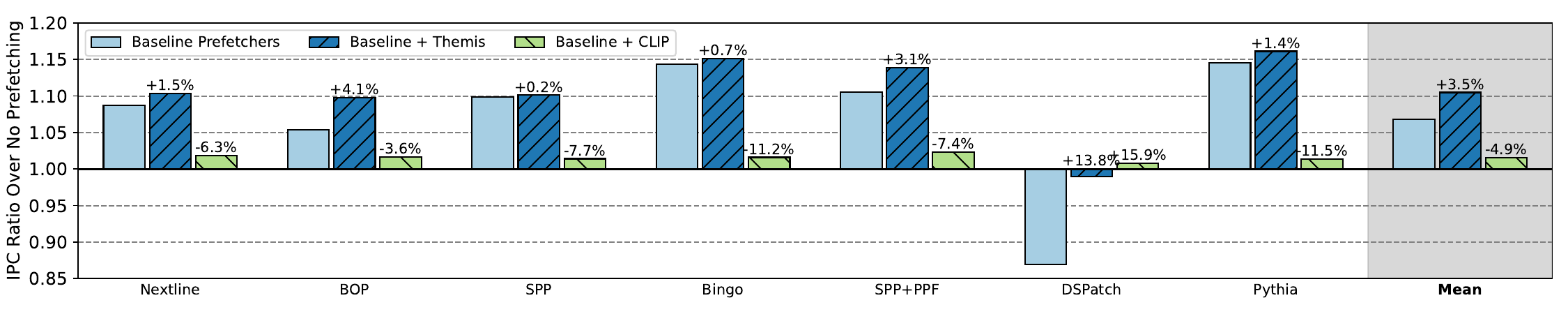}
    \caption{IPC ratios of baseline prefetchers, \sys, and CLIP over no prefetching for Google traces. The numbers above each bar show the speedup from the baseline prefetchers (leftmost bars).}
    \label{fig:eval-main-result}
\end{figure*}

\begin{table}[t]
    \centering
    \caption{Simulation parameters}
    \begin{tabular}{p{0.25\linewidth}p{0.7\linewidth}}
    \hline
        \textbf{Parameter} & \textbf{Value} \\\hline
        CPU & $6$-wide OoO, $352$-entry ROB, $128$/$72$-entry LQ/SQ\\\hline
        Branch pred. & Bimodal \\\hline
        L1/L2 Caches & $48\,\mathrm{KB}$/$512\,\mathrm{KB}$, $64\,\mathrm{B}$ line, 12/8 way, LRU, 16/32 MSHRs, 5-cycle/10-cycle latency \\\hline
        LLC & $2\,\mathrm{MB}$, $64\,\mathrm{B}$ line, 16 way, LRU, 64 MSHRs, 20-cycle latency\\\hline
        Main Memory & Single channel, 1 rank per channel, 8 banks per rank, $3200\,\mathrm{MTPS}$, $64$-bit data bus per channel, $64\,\mathrm{KB}$ row buffer per bank, tRP=$12.5\,\mathrm{ns}$, tRCD=$12.5\,\mathrm{ns}$, tCAS=$12.5\,\mathrm{ns}$ ($25.6\,\mathrm{GB/s}$)\\\hline 
    \end{tabular}
    \label{tab:simulation_parameters}
\end{table}

\section{Evaluation}
\label{sec:eval}

\subsection{Methodology}
\label{sec:eval-method}
\pgheading{Workloads}
For the evaluation, we utilize Google datacenter application traces (version 2)~\cite{Google_Workload_Traces_Version_2}. These traces include instruction and memory address traces, and while architecture-specific details such as instruction opcodes are omitted, they provide enough information about register dependencies such that they can be realistically simulated in ChampSim.

Because Google traces are multi-threaded, we transform the original per-thread traces into per-core traces by leveraging DynamoRIO’s thread scheduling mechanism~\cite{dynamorio}, as recommended by Google. We employ the default random scheduling strategy to assign threads to cores. For each workload, we assume that it is running on the number of cores specified by Google as the workload's peak core limit. We convert the resulting per-core traces, which include interleaved execution from multiple threads, to a ChampSim compatible trace format. Google traces contain more than 40 billion instructions per application on average.

To ensure that we simulate representative workload regions, for each workload, we randomly sample ten distinct regions of 100 million instructions across all the cores. We report the results for each workload by averaging statistics from all ten workload samples.

We also evaluate \sys on common benchmarks, including SPEC2017~\cite{SPEC2017} and GAP benchmark suites~\cite{beamer2015gap}, using traces provided by DPC-3~\cite{DPC3} and the 1st championship value prediction (CVP-1)~\cite{CVP1}, respectively.
We only consider traces that have more than \mpkithresh last-level cache misses per kilo instruction (MPKI) when prefetching is disabled.

\pgheading{Profiling Methodology}
To model the profiling phase of \sys, we profile a single 100-million-instruction region from a single core for each workload. The collected profile data is then shared across all ten simulated samples for that workload. We ensure a gap of at least 100 million instructions between the profiling region and the regions used for performance measurement. In other words, one profile data from a 100-million-instruction region is tested across multiple program regions sampled from tens of billions of instructions.
In the experiments, we assume prefetchers are working on physical addresses; hence, there is no additional overhead to look up the page attribute values.
Unless mentioned otherwise, we show data for Google traces.
Table~\ref{tab:simulation_parameters} summarizes the configurations of the simulated CPU.

\subsection{Performance Analysis}
\label{sec:eval-perf}
\pgheading{IPC Improvement}
Figure~\ref{fig:eval-main-result} shows the IPC improvement of \sys and CLIP for Google traces.
For each prefetcher, the geometric mean of IPC improvement over no prefetching is shown for baseline prefetchers, \sys, and CLIP.
The numbers above each bar show the speedup ratio brought by using \sys or CLIP against baseline prefetchers.

Overall, \sys successfully improves the IPC over the baseline for all the prefetchers. 
Across seven prefetchers, geometric mean of speedup is $3.5\%$ ($0.4\%$ for \texttt{arizona}, $3.1\%$ for \texttt{bravo.a}, $0.6\%$ for \texttt{charlie}, $2.8\%$ for \texttt{delta}, $2.2\%$ for \texttt{merced}, $4.7\%$ for \texttt{sierra.a.3}, $5.6\%$ for \texttt{sierra.a.4}, $8.1\%$ for \texttt{sierra.a.6}, $3.4\%$ for \texttt{tahoe}, $4.2\%$ for \texttt{tango}, $2.7\%$ for \texttt{whiskey}, and $3.9\%$ for \texttt{yankee}).
Pythia initially delivered the best performance, and \sys further enhances this by an additional $1.4\%$.
In contrast, CLIP results in an average slowdown of $4.9\%$, showing its ineffectiveness for datacenter workloads.
Although CLIP outperforms \sys in the case of DSPatch, the baseline performance for DSPatch is lower than that of the no-prefetch configuration, and the speedup relative to no-prefetch is minimal (less than $1\%$) for both \sys and CLIP. The DSPatch baseline emits a lot of incorrect prefetches as it cannot handle the large data footprints. We will show in \S\ref{sec:eval-sensitivity} that DSPatch requires large metadata storage to deliver adequate performance without throttling.

\begin{figure}[t]
    \centering
    \includegraphics[width=\columnwidth]{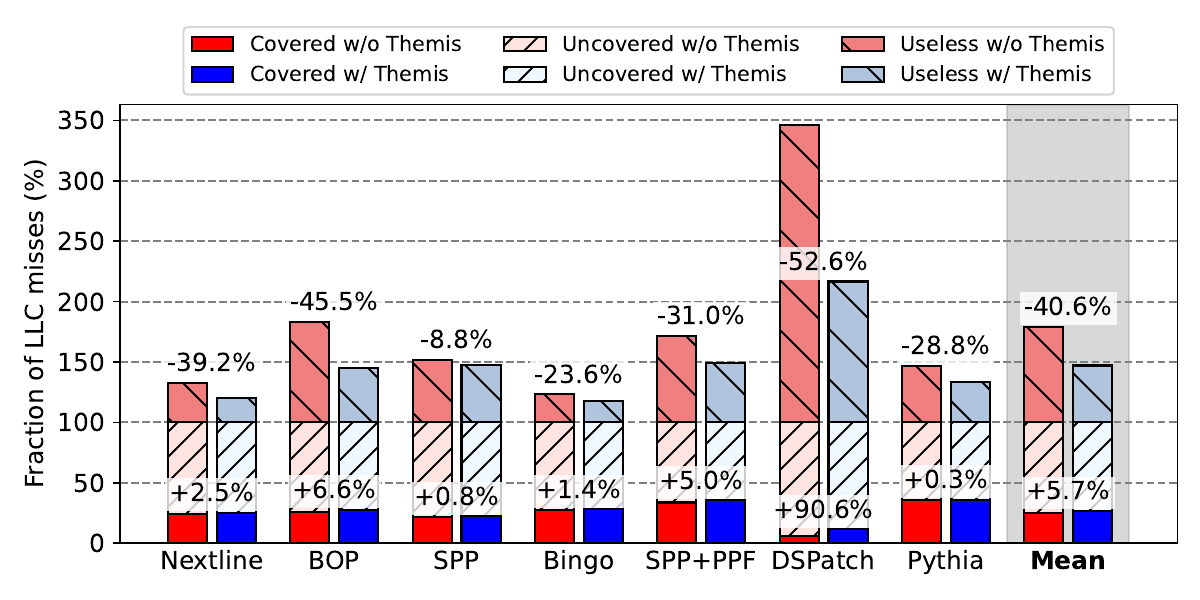}
    \caption{Coverage and useless prefetches of each prefetcher with and without \sys. Numbers show the reduction ratio of useless and useful prefetches.}
    \label{fig:eval-coverage}
\end{figure}

\pgheading{Prefetch Accuracy and Coverage}
The performance improvement of \sys comes from reducing useless prefetches while preserving useful prefetches.
Figure~\ref{fig:eval-coverage} compares the breakdown of LLC misses with and without \sys in a similar way as Figure~\ref{fig:analysis-prefetcher-performance}~(b), showing the average across all of the workloads of Google traces.
The numbers above and within each bar show reduction ratios of useless prefetches and useful prefetches, respectively.

We can observe that \sys reduces the number of useless prefetches significantly. 
The reduction is most significant for DSPatch, decreasing by $52.6\%$.
On average, $40.6\%$ of useless prefetches are eliminated due to \sys.
Even in the worst case (SPP), the useless prefetches are reduced by $8.8\%$.
Thanks to this reduction, the overall DRAM bandwidth utilization is decreased by $18.0\%$ on average.

Moreover, there is a slight increase in coverage (\ie the number of useful prefetches).
Notably, DSPatch exhibits the largest gain, with a $90.6\%$ increase in coverage, which contributes to an average improvement of $5.7\%$. Even in the least favorable scenario, as seen with Pythia, coverage still increases by $0.3\%$.

Although \sys functions solely as a throttling mechanism without directly issuing additional prefetches, it enhances overall coverage. It does so by optimizing hardware utilization during training and mitigating cache pollution. Specifically, \sys can disable the training on instruction streams where all data pages are marked as disabled.
For prefetchers that learn spatial access patterns within pages, this means that resources are not wasted storing information for pages where prefetches are unlikely to be effective. As a result, the limited hardware storage capacity is concentrated on regions where useful prefetches can be issued, thereby improving the effectiveness of the hardware training resources and increasing coverage.

Moreover, by reducing cache pollution, the overall number of cache misses is decreased, which contributes to a higher coverage value. For prefetchers like DSPatch that adjust their prefetch aggressiveness based on bandwidth utilization, disabling prefetches for certain pages allows these prefetchers to allocate more resources to enabled pages, further boosting overall coverage.

\begin{figure}[t]
    \centering
    \includegraphics[width=\columnwidth]{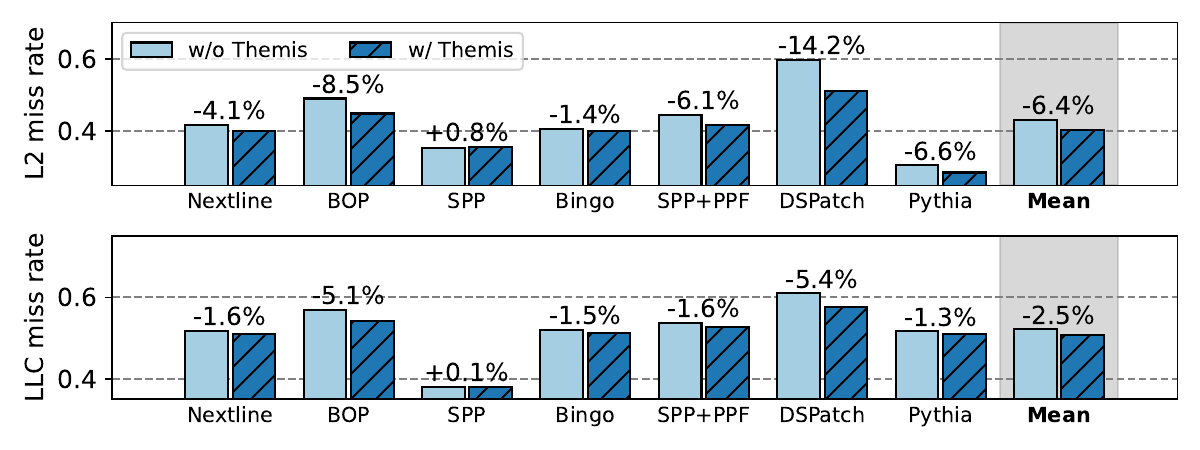}
    \caption{Cache miss rates with and without \sys for L2 cache and LLC for Google traces.}
    \label{fig:eval-miss-rate}
\end{figure}

\pgheading{Cache Miss Rate}
Figure~\ref{fig:eval-miss-rate} shows the average miss rate of L2 cache and LLC for Google traces with and without \sys. 
We observe a reduction in the LLC and L2 cache miss rate for all the prefetchers except for SPP, where the increase in miss rate is less than $1\%$. 
On average, \sys reduces L2 misses by $6.4\%$ and LLC misses by $2.5\%$.
This data suggests that \sys effectively reduces cache pollution by eliminating useless prefetches.

\begin{figure}[t]
    \centering
    \includegraphics[width=\columnwidth]{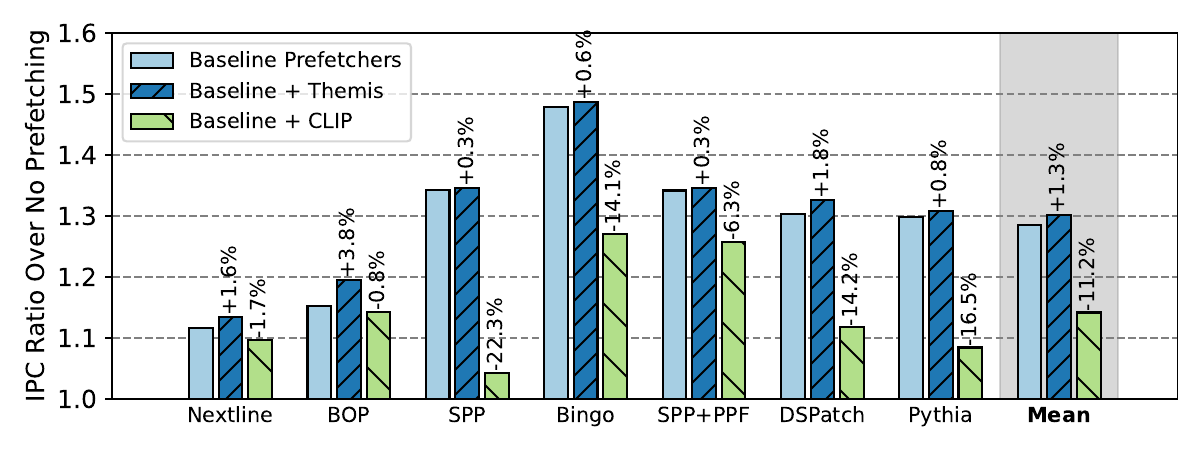}
    \caption{IPC of baseline prefetchers, \sys, and CLIP for SPEC2017 benchmark suite. Numbers above bars show the speedup from the baseline prefetchers (leftmost bars).}
    \label{fig:eval-ipc-spec}
\end{figure}

\pgheading{SPEC2017}
We also evaluate \sys on conventional applications to show its effectiveness beyond datacenter applications.
Figure~\ref{fig:eval-ipc-spec} shows the performance of baseline prefetchers, \sys, and CLIP for SPEC2017 benchmarks.

Similar to the case of Google traces, we observe positive performance improvement for all the prefetchers by using \sys.
\sys also outperforms CLIP in all the cases, which fails to improve the IPC since it is optimized for more bandwidth-constrained environments.
However, the ratio of acceleration is smaller than for Google traces.
\sys improves the IPC by $3.5\%$ on a geometric mean for Google traces, while $1.3\%$ for SPEC2017.
This result shows that \sys is especially suited for optimizing the performance of datacenter applications with large instruction and memory footprints.

\begin{figure}[t]
    \centering
    \includegraphics[width=\columnwidth]{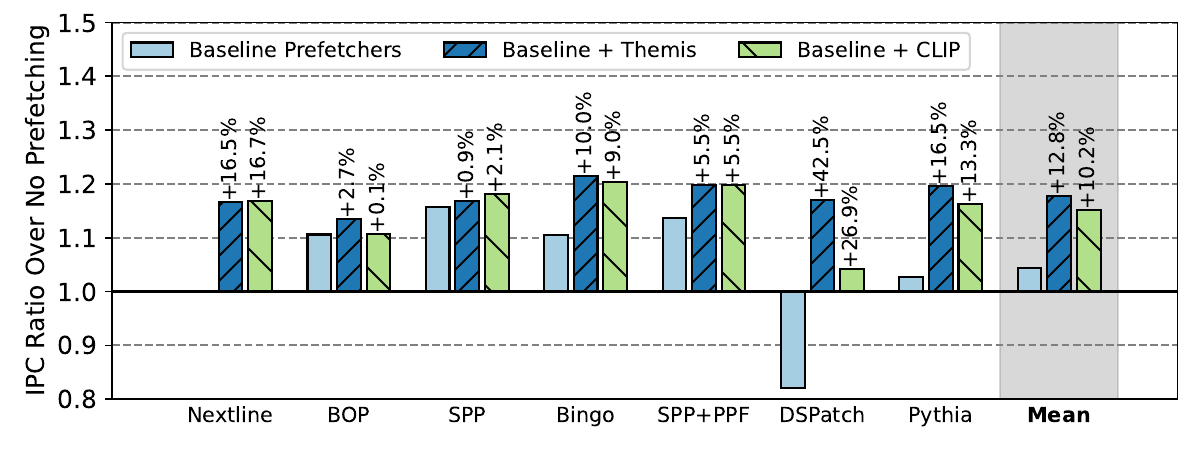}
    \caption{IPC of baseline prefetchers, \sys, and CLIP for GAP benchmark suite. Numbers above bars show the speedup from the baseline prefetchers (leftmost bars).}
    \label{fig:eval-ipc-gap}
\end{figure}

\pgheading{GAP}
Figure~\ref{fig:eval-ipc-gap} shows the performance for GAP benchmark suite.
Again, the performance of baseline prefetchers, \sys, and CLIP are compared.
GAP consists of memory-intensive applications and benefits significantly from prefetch throttling, as both \sys and CLIP improve IPC for all the prefetchers.
On a geometric mean, \sys increases IPC by $12.8\%$, which outperforms CLIP's performance ($10.2\%$ increase of IPC).
CLIP performs better than \sys for SPP, but for the other six out of seven prefetchers, \sys's provided speedups are superior.

\begin{figure}[t]
    \centering
    \includegraphics[width=\columnwidth]{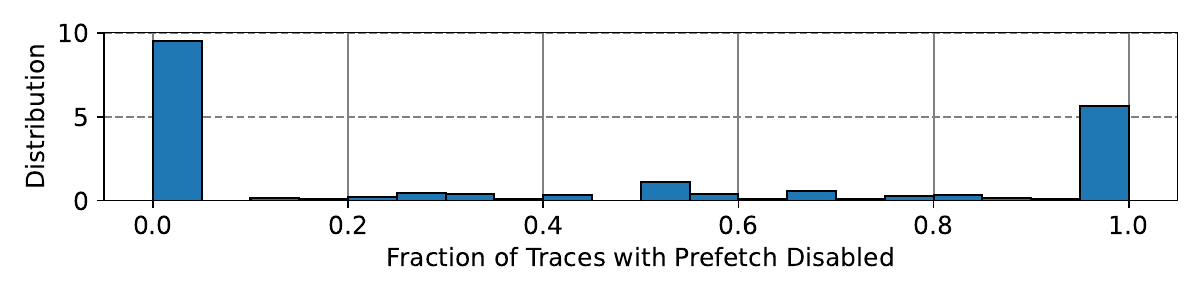}
    \caption{Histogram showing how often each page has prefetching disabled.}
    \label{fig:eval-prefetch-consistency}
\end{figure}

\subsection{Overhead Analysis}
\label{sec:eval-overhead}

Next, we quantify the profiling overhead of \sys by measuring two metrics: (1) how often does the profiler need to run, and (2) what is the overhead of the profiling phase?

\pgheading{Profiling Frequency}
To demonstrate that frequent profiling is not necessary for \sys, we examine the consistency of prefetch usefulness data across different trace segments from the same workload---the segments span tens of billions of instructions in the workload.

For each workload, we collect profiling data from ten distinct traces and then analyze how often page addresses were marked to disable prefetching. Figure~\ref{fig:eval-prefetch-consistency} presents a histogram as follows: we divide the number of traces that disabled prefetching by the total number of traces in which the page appears. This provides a single float value between 0 and 1 for each page, and the histogram shows the overall density distribution of these values. We only include pages that appear in at least two traces. The histogram shows that most pages have ratios close to 0 or 1, indicating that the decision to disable prefetching is very consistent for individual pages across different trace segments sampled from tens of billions of instructions. 
Thus, \sys only profiles once per at least 40 billion instructions or once per around $25\,\mathrm{s}$ of execution time (assuming an IPC of 0.6 and frequency of $3\,\mathrm{GHz}$).

\begin{figure}[t]
    \centering
    \includegraphics[width=\columnwidth]{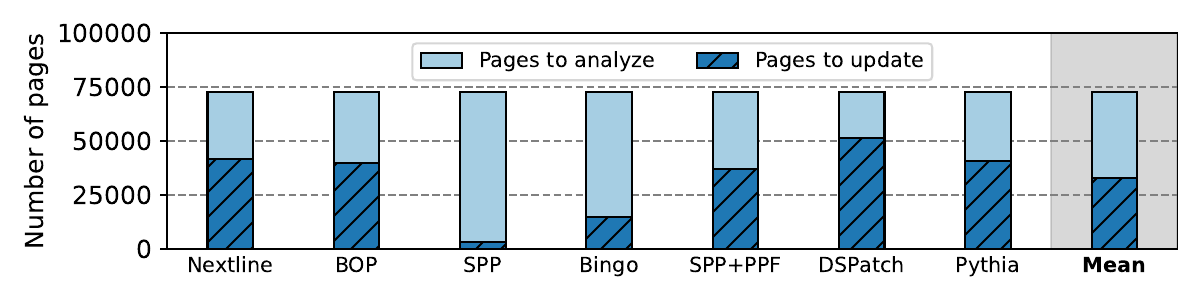}
    \caption{The number of pages to be analyzed and the PTEs to be updated for \sys with 100 million instructions profile for Google traces.}
    \label{fig:eval-overhead-analysis}
\end{figure}

\pgheading{Page-Table Updates}
The profiling overhead of \sys comprises two parts: (1) computing Equation~\ref{eq:condition} for each page and (2) updating the page attributes.
Figure~\ref{fig:eval-overhead-analysis} shows the number of times each operation is required for a profiling window of 100 million instructions.
Over 100 million instructions, the profiler needs to analyze 73000 pages, resulting in 33000 page-table attribute updates. Note that these updates are only during the initial profiling phase, when no directives were initialized. Iterative profiling, where profiles are updated at fixed intervals, will require much fewer updates.

As discussed in \S\ref{sec:design-overhead}, updating PTE attributes takes approximately $1\,\mu\mathrm{s}$ per bit (including the cost of page-table walk). For a profile consisting of 100 million instructions and 33000 updates, the total latency required for updating the page-table is ~$33\,\mathrm{ms}$. Since a single profiling dataset remains effective for multiple billions of instructions, as demonstrated by our experimental results, the overhead of PTE updates can be effectively amortized over long execution times.
More concretely, if the profiler runs once every $15\,\mathrm{s}$ (tens of billions of instructions), and it adds $33\,\mathrm{ms}$ of execution time, this results in $0.2\%$ execution overhead for the profiler.

\begin{figure}[t]
    \centering
    \includegraphics[width=\columnwidth]{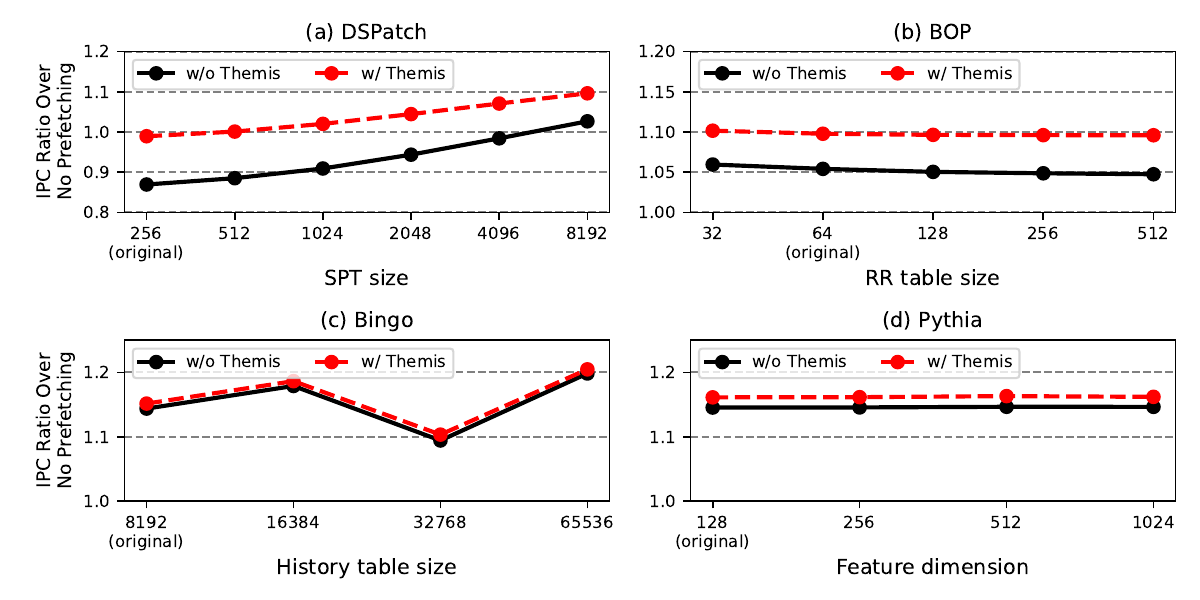}
    \caption{IPC with different sizes of metadata storage.}
    \label{fig:eval-storage}
\end{figure}

\subsection{Sensitivity Analysis}
\label{sec:eval-sensitivity}

\pgheading{Storage Overhead for Hardware Prefetchers}
As \sys disables the training of hardware prefetchers for certain pages, it reduces the pressure on the metadata storage of prefetchers.
As a result, prefetchers can better utilize the limited hardware resources.
To show this effect, we evaluate some of the prefetchers by varying their storage size.
Figure~\ref{fig:eval-storage} shows the performance with and without \sys with varying sizes of on-chip resources for (a) DSPatch, (b) BOP, (c) Bingo, and (d) Pythia.
We vary SPT size for DSPatch, recent requests (RR) table size for BOP, history table size for Bingo, and feature dimension for Pythia.

For DSPatch, there is a clear positive correlation between the SPT size and IPC. 
Using \sys on the baseline has a similar effect as increasing the SPT size by $32$ times (from $256$ to $8192$).
BOP's performance is not highly correlated with the RR table size, and using \sys has a more significant effect than increasing the buffer size.
For Bingo, IPC is positively correlated with history table size except for the case of $32768$, and using \sys consistently increases the IPC.
For Pythia, \sys has a more significant effect on the performance than increasing the feature dimension.
Overall, using \sys has a similar effect as increasing the buffer size (DSPatch) or a more significant effect than that (BOP, Pythia).

\begin{figure}[t]
    \centering
    \includegraphics[width=\columnwidth]{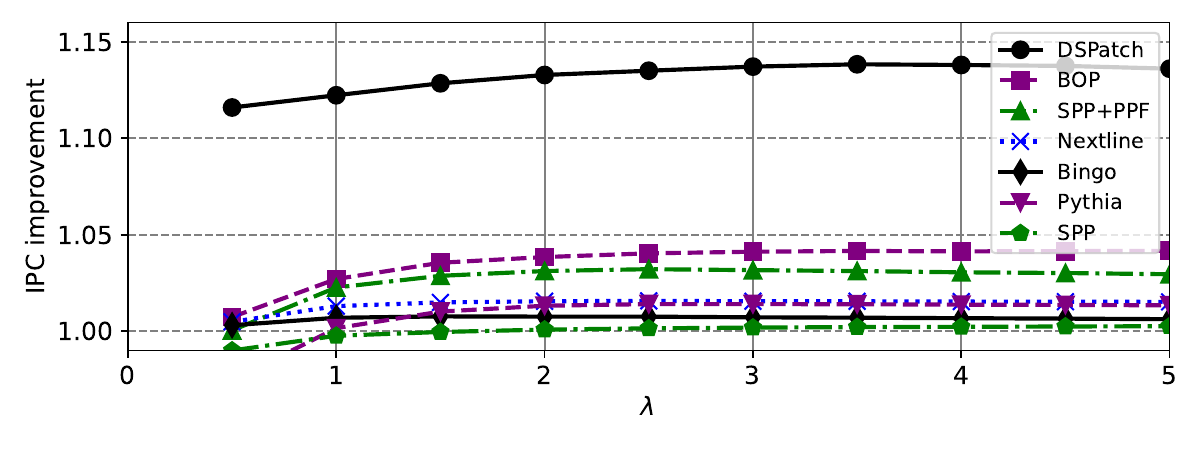}
    \caption{IPC improvement of \sys over baseline prefetchers as a function of \lambdaname $\lambda$.}
    \label{fig:eval-lambda-sensitivity}
\end{figure}

\pgheading{Value of $\lambda$}
To show the effect of \lambdaname $\lambda$ to \sys, we evaluate the performance by varying the value of $\lambda$.
Figure~\ref{fig:eval-lambda-sensitivity} shows the sensitivity of IPC with \sys with different values of $\lambda$ for each prefetcher.
As the value of $\lambda$ decreases, the throttling by \sys becomes stronger, and fewer prefetches will be issued.
The performance gets better as $\lambda$ increases from $0$ and reaches the peak at $4$ on average.
However, the performance is not very sensitive to $\lambda$.
For example, from $\lambda = 2.5$ to $5$, the difference between the best and worst IPC improvements is less than $1\%$ on average.

\begin{figure}[t]
    \centering
    \includegraphics[width=\columnwidth]{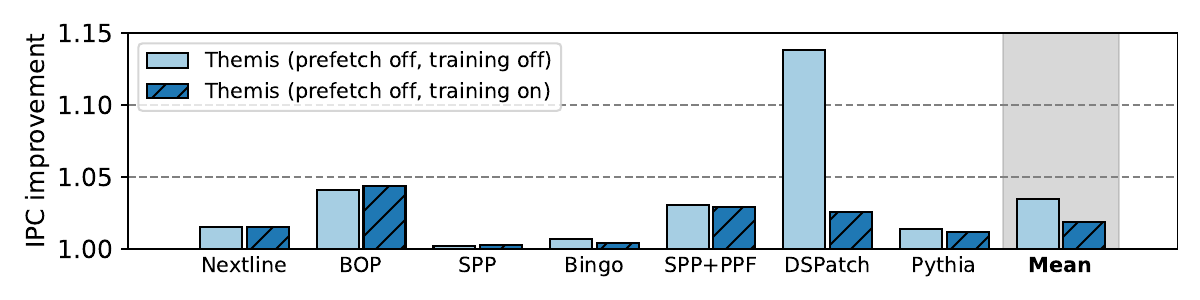}
    \caption{IPC improvement over baseline prefetchers of \sys and a variant of \sys that does not disable training.}
    \label{fig:eval-visible}
\end{figure}

\pgheading{Training on Disabled Pages}
With \sys, prefetchers are not trained on memory access patterns for pages where prefetching is disabled, so they can efficiently utilize their hardware resources.
Therefore, if we disable the prefetching but do not disable the training, we expect to get lower performance as it does not lead to the efficient utilization of hardware resources.

To show the effect of disabled training, Figure~\ref{fig:eval-visible} compares the IPC improvement of \sys against a variant of \sys that does not disable the training of prefetchers.
As expected, we observe less performance improvement if the training is not disabled for all the prefetchers.
On average, the training-enabled version has only $1.9\%$ of improvement while the proposed version improves by $3.5\%$.
DSPatch shows the biggest difference, where the speedup is $13.8\%$ and $2.6\%$ for the original and training-enabled versions, respectively.

\begin{figure}[t]
    \centering
    \includegraphics[width=\columnwidth]{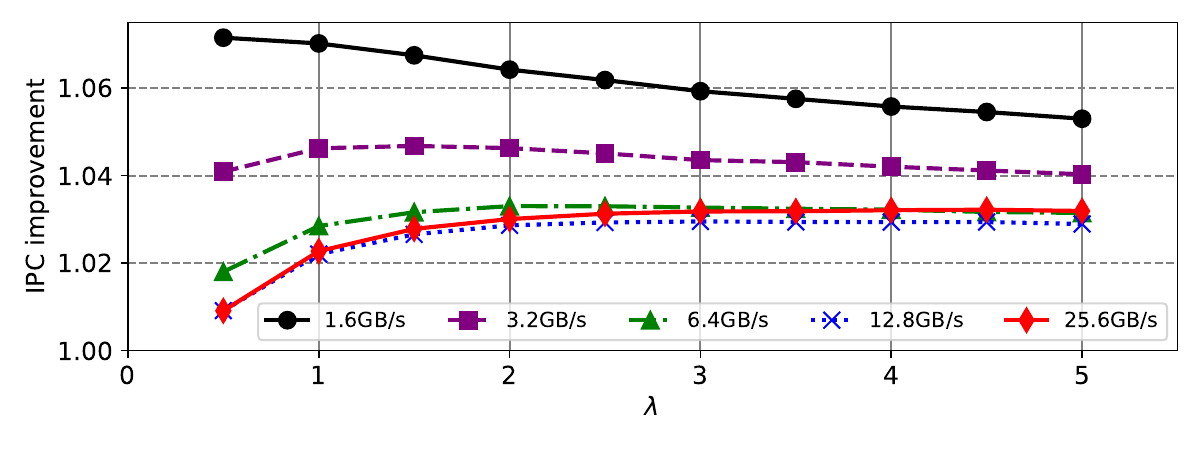}
    \caption{Average IPC improvement of \sys with different DRAM bandwidths as a function of \lambdaname $\lambda$.}
    \label{fig:eval-bandwidth-sensitivity}
\end{figure}

\pgheading{DRAM Bandwidth}
To evaluate \sys's performance in environments with more constrained bandwidth, which is prevalent in many-core architectures, we simulate the system by scaling the DRAM's million transfers per second (MTPS).
Figure~\ref{fig:eval-bandwidth-sensitivity} shows the performance of \sys with different DRAM bandwidths.
The graph shows the geometric mean of IPC improvement across seven prefetchers as a function of \lambdaname $\lambda$\footnote{Here, we show the results with one trace per workload due to long simulation time.}.

Overall, the effect of \sys improves as the available DRAM bandwidth decreases.
Moreover, the optimal value of $\lambda$ shifts as the bandwidth changes.
In the case of $25.6\,\mathrm{GB/s}$ ($3200\,\mathrm{MTPS}$), $\lambda = 4$ works best.
On the other hand, the optimal value of $\lambda$ is $3$ for $12.8\,\mathrm{GB/s}$ ($1600\,\mathrm{MTPS}$), $2$ for $6.4\,\mathrm{GB/s}$ ($800\,\mathrm{MTPS}$), $1.5$ for $3.2\,\mathrm{GB/s}$ ($400\,\mathrm{MTPS}$), and $0.5$ for $1.6\,\mathrm{GB/s}$ ($200\,\mathrm{MTPS}$), meaning stronger throttling is desirable when the bandwidth is scarce.
This is because the negative effect of useless prefetches is more significant when less bandwidth is available, consistent with previous works~\cite{panda2023clip}.

\begin{figure}[t]
    \centering
    \includegraphics[width=\columnwidth]{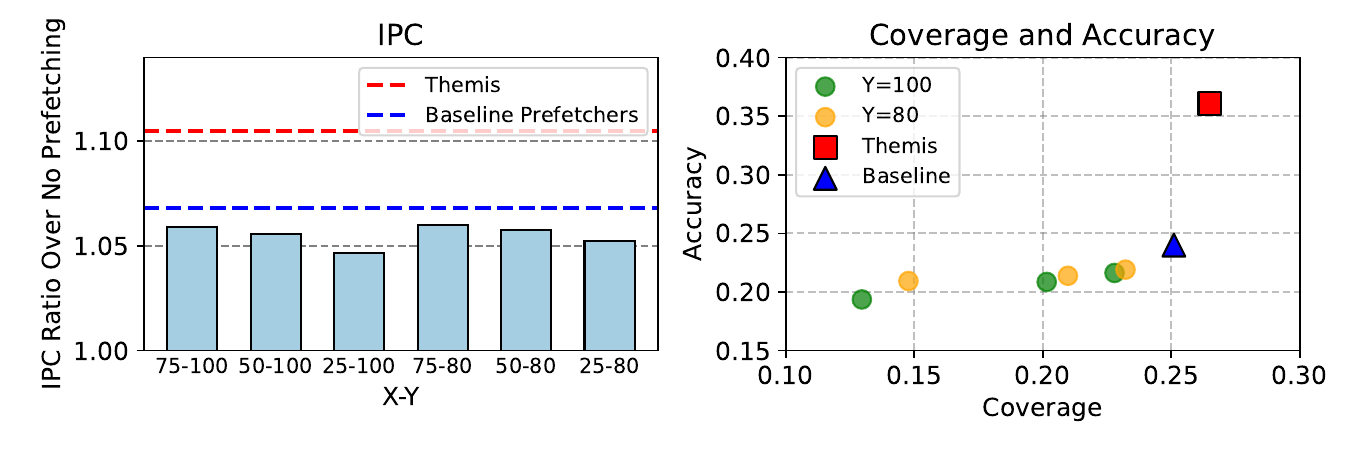}
    \caption{Performance with simple accuracy- and bandwidth-based prefetch throttling.}
    \label{fig:eval-cbusy}
\end{figure}

\begin{figure}[t]
    \centering
    \includegraphics[width=\columnwidth]{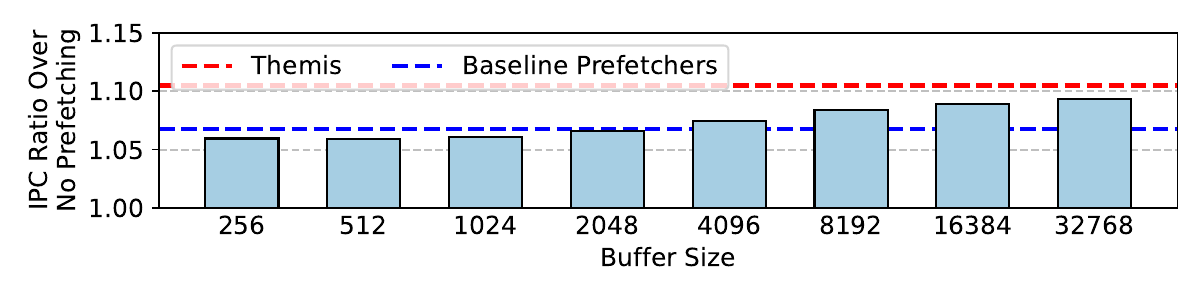}
    \caption{Performance of hardware-only throttling based on page addresses with different buffer sizes (defined as the number of pages to track).}
    \label{fig:eval-hw-only}
\end{figure}

\pgheading{Hardware-Only Baselines}
To demonstrate that hardware-software collaboration is necessary for \sys, Figure~\ref{fig:eval-cbusy} evaluates hardware-only throttling mechanisms similar to prior work~\cite{arm2025useOfCBusy,srinath2007feedback}. These approaches disable prefetching when memory bandwidth utilization exceeds $X\%$ and prefetch accuracy falls below $Y\%$ with varied $X$ and $Y$. However, none of the evaluated configurations outperform baseline prefetchers across IPC, accuracy, and coverage, demonstrating that coarse-grained throttling is insufficient for effective prefetch management in datacenter workloads.

Moreover, Figure~\ref{fig:eval-hw-only} shows a hardware-only throttling mechanism that tracks prefetch usefulness per page using an LRU buffer storing page addresses and useful/useless counters. Even with a $32\mathrm{k}$ buffer size, this approach achieves lower IPC than \sys while requiring $128\,\mathrm{KB}$ of storage (assuming $32$\,bits per entry to store page address and prefetch useful/useless counts).
This shows hardware-only mechanisms require prohibitive storage overhead to match \sys, highlighting the necessity of software assistance.

\section{Related Work}
\label{sec:related}
\pgheading{Hardware Prefetching}
Existing hardware prefetchers can be roughly divided into spatial and temporal prefetchers.
Spatial prefetchers learn memory access patterns over spatial memory regions and predict the memory access as an offset from the base address or spatial bit pattern~\cite{baer1991effective,chen1995effective,fu1992stride,ishii2009access,jouppi1990improving,kondguli2018division,kumar1998exploiting,pugsley2014sandbox,shakerinava2019multi,somogyi2006spatial,srinath2007feedback,smith1978sequential,michaud2016best,kim2016path,bakhshalipour2019bingo,bhatia2019perceptron,bera2019dspatch,pakalapati2020bouquet,bera2021pythia,navarro2022berti,gerogiannis2023micro}.
They achieve higher accuracy with smaller storage overhead.

Temporal prefetchers learn the sequence of memory addresses accessed by the application~\cite{bakhshalipour2018domino,bekerman1999correlated,chilimbi2002dynamic,chou2007low,cooksey2002stateless,ferdman2007last,hu2003tcp,jain2013linearizing,joseph1997prefetching,karlsson2000prefetching,somogyi2009spatio,wenisch2009practical,wenisch2010making,wenisch2005temporal,wu2019temporal,wu2019efficient,ainsworth2024triangel}. 
When a cache line is accessed again in the future, temporal prefetchers issue prefetches to addresses that were accessed following the cache line in the previous access.
This type of prefetcher usually requires a larger amount of metadata storage than spatial prefetchers and often stores metadata in memory.

Several works attempt to use machine learning-based methods to learn access patterns~\cite{hashemi2018learning,peled2015semantic,shi2019neural,shi2021hierarchical,shi2019learning,teran2016perceptron,bera2021pythia,gerogiannis2023micro,duong2024new,jia2024pathfinder,fu2024differential}.
PPM~\cite{vavouliotis2022page} uses page size information to detect access patterns over page boundaries, and MOKA~\cite{vavouliotis2025cross} is a framework for cross-page prefetching incorporating hashed perceptron predictors.
Hermes~\cite{bera2022hermes} is a lightweight perceptron-based predictor for off-chip load requests.

\sys is orthogonal to these works.
By using \sys with these prefetchers, we can reduce useless prefetches and achieve more efficient utilization of hardware resources, improving the overall performance.

Another line of research focuses on precomputation-based prefetching.
They generate accurate prefetches either by runahead execution~\cite{dundas1997improving,hashemi2016continuous,hashemi2015filtered,iacobovici2004effective,mutlu2005address,mutlu2005techniques,mutlu2006efficient,mutlu2005reusing,mutlu2003runahead,naithani2021vector} or helper thread execution~\cite{chappell1999simultaneous,collins2001dynamic,collins2001speculative,jung2006helper,kadjo2014b,luk2001tolerating,ramirez2008runahead,solihin2002using,wang2004helper,zhang2007accelerating,zilles2001execution}, but they have high overhead.
In contrast, \sys uses a concurrent thread just for analyzing profile data and the overhead is small as discussed in \S\ref{sec:eval-overhead}.

\pgheading{Software Prefetching} 
Early software prefetching~\cite{callahan1991software} used static induction variable analysis to prefetch stride patterns inside loops.~\citet{mowry1992design} extended it by identifying highly reused addresses to limit prefetching, and using loop splitting to prefetch ahead without branch conditions.~\citet{ainsworth2017software} proposed automatic compiler-injected prefetching for more complex irregular accesses. APT-GET~\cite{jamilan2022apt} improved timeliness of these prefetches by profiling Intel PMU counters. 

Software prefetching methods have promise in niche memory access patterns like irregular accesses. They could work in tandem with \sys, which focuses on improving hardware prefetcher accuracy for \emph{all} prefetches, that software prefetching may not cover.

\pgheading{Prefetch Throttling}
Several methods are proposed to throttle hardware prefetchers to get higher accuracy and save memory bandwidth usage~\cite{panda2016spac,srinath2007feedback,ebrahimi2009coordinated,heirman2018near,navarro2020bandwidth}. Unlike them, \sys coordinates with software to enable fine-grained throttling without requiring a large amount of storage.
A state-of-the-art prefetch throttler, CLIP~\cite{panda2023clip}, blocks prefetches based on the criticality of the load instruction (\ie if it leads to re-order buffer stalls) and the accuracy of prefetches.
Unlike it, \sys uses region-based control and overcomes the limitation of hardware resources.
Limoncello~\cite{jain2024limoncello} dynamically switches between hardware and software prefetching based on real-time memory bandwidth telemetry for datacenters.
TLP~\cite{jamet2024two} proposes the two-level perceptron to optimize the memory subsystem by combining off-chip memory access prediction and adaptive prefetch filtering.

\pgheading{Profile-Guided Optimization (PGO)}
PGO is a method that optimizes CPU performance by using profile information of the target workload~\cite{panchenko2021lightning,ottoni2021hhvm}. 
In the past, PGO has been used to optimize various components of microarchitecture such as instruction cache prefetching~\cite{khan2020spy} and replacement~\cite{khan2021ripple}, BTB prefetching~\cite{khan2021twig} and replacement~\cite{song2022thermometer}, branch predictor~\cite{khan2022whisper}, and software prefetching~\cite{jamilan2022apt,ayers2020classifying,litz2022crisp}, but these methods have another set of disadvantages~\cite{zhang2022ocolos,zhang2023online}.
\sys is a throttling technique for hardware data prefetcher based on the directives by the software that works online and overcomes the disadvantages of existing PGO techniques.

\pgheading{Hardware-Software Co-Design} 
Guided-region prefetching (GRP)~\cite{wang2003guided} uses compiler static-analysis to detect loop bounds of load instructions. This is used to guide the underlying region prefetcher~\cite{lin2001reducing} to limit the size of prefetching region and reduce traffic overhead. GRP, unlike \sys, relies on static analysis which can only find a small subset of strided instructions in large scale workloads with limited function inlining and complex control flow.
Efficient content-directed prefetching (ECDP)~\cite{ebrahimi2009techniques} uses profile-guided hints to score all possible prefetch offsets in a physical page, for each pointer-based instruction. ECDP focuses on a domain-specific prefetcher~\cite{cooksey2002stateless} for linked data structures (LDS), and requires a finer-grained one-hot encoded 16-bit hint for each instruction. \sys is lightweight and versatile across state-of-the-art prefetchers.

\section{Conclusion}
\label{sec:conclusion}

We introduce \sys, a hardware-software co-designed technique to optimize prefetchers based on the profile information. \sys dramatically improves the efficiency of hardware prefetchers without compromising their coverage. We accomplish this by disabling the hardware prefetcher for certain data pages based on runtime profile information. \sys is orthogonal to existing work on hardware prefetching, and it can be applied to further optimize any hardware prefetcher. Furthermore, it requires no binary modification and no ISA changes. 
Our evaluation on datacenter applications demonstrates positive speedup for all the evaluated prefetchers, including $4.1\%$ for BOP, $3.1\%$ for SPP+PPF, and $1.4\%$ for Pythia.

\bibliographystyle{ACM-Reference-Format}
\bibliography{_refs}

\end{document}